\documentclass{article}
\usepackage{amsfonts}
\usepackage{spconf,amsmath,graphicx,epsfig}
\usepackage{subcaption}
\usepackage{mwe}
\usepackage{xcolor}
\usepackage{hyperref}
\usepackage{algorithm}
\usepackage{algorithmic}
\usepackage{multirow}
\usepackage{braket}
\usepackage{qcircuit}
\usepackage{adjustbox}
\usepackage{booktabs}
\usepackage{mathtools}

\usepackage{color}

\title{From English to More Languages: Parameter-Efficient Model Reprogramming for Cross-Lingual Speech Recognition}

\name{%
\begin{tabular}{@{}c@{}}
Chao-Han Huck Yang\thanks{$^{*}$Work done during a research internship at Google.}$^{*1,2}$\qquad Bo Li$^{1}$\qquad Yu Zhang$^{1}$\qquad Nanxin Chen$^{1}$ \\ Rohit Prabhavalkar$^{1}$\qquad Tara N. Sainath$^{1}$\qquad Trevor Strohman$^{1}$
\end{tabular}}
\address{$^1$Google, USA \\$^2$Georgia Institute of Technology, USA}

\begin{document}
\ninept
\maketitle
\begin{abstract}
In this work, we propose a new parameter-efficient learning framework based on neural model reprogramming for cross-lingual speech recognition, which can \textbf{re-purpose} well-trained English automatic speech recognition (ASR) models to recognize the other languages.  We design different auxiliary neural architectures focusing on learnable pre-trained feature enhancement that, for the first time, empowers model reprogramming on ASR. Specifically, we  investigate how to select trainable components (i.e., encoder) of a conformer-based RNN-Transducer, as a frozen pre-trained backbone. Experiments on a seven-language multilingual LibriSpeech speech (MLS) task show that model reprogramming only requires $4.2$\% (11M out of 270M) to $6.8$\% (45M out of 660M) of its original trainable parameters from a full ASR model to perform competitive results in a range of $11.9$\% to $8.1$\% WER averaged across different languages. In addition, we discover different setups to make large-scale pre-trained ASR succeed in both monolingual and multilingual speech recognition. Our methods outperform existing ASR tuning architectures and their extension with self-supervised losses (e.g., w2v-bert) in terms of lower WER and better training efficiency.

\end{abstract}

\section{Introduction}
\label{sec1}
Recent advances~\cite{he2019streaming, sainath2015deep, chiu2018state, li2021scaling, li2022massively, li2020towards} in developing large-scale ASR architectures have demonstrated promising results for English speech recognition tasks. Moreover, English ASR model with self-supervised training objectives, such as wav2vec2~\cite{baevski2020wav2vec}, w2v-BERT~\cite{chung2021w2v}, and BigSSL~\cite{zhang2022bigssl},  further boosts recognition performance, as an extension from the existing supervised ASR framework with annotated data. Meanwhile, the success of current neural ASR models is \textbf{still related to the scale} of training data, where training large neural ASR models does not always ensure competitive results in medium or small-scale corpora for non-English and low resource languages conditions. When current ASR data are mainly based on English~\cite{zhang2022bigssl}, how to advance the power of well-trained English ASR systems (e.g., RNN-T~\cite{graves2012sequence}) to \emph{other languages}~\cite{hu2019phoneme} is still an open question that could \textbf{benefit more worldwide end-users}.

Previous works on pre-training and fine-tuning encoders of English ASR models have demonstrated some recent success in 
the West Germanic languages~\cite{tong2017investigation, bansal2019pre} (e.g., English and Deutsch), atypical~\cite{fan2022draft}, and accent speech recognition~\cite{tomanek2021residual}. Motivated by the previous discussion, we aim to investigate how to efficiently transfer large-scale English ASR for both monolingual and multilingual speech recognition in this work. One notorious challenge of applying large-scale ASR for mobile applications is the model complexity (e.g., trainable parameters) in terms of memory. Tuning a large-scale ASR for a new task or dataset often requires a significant training cost (e.g., time and power), which further makes large pre-trained speech models difficult to deploy on mobile and smart home voice applications in terms of latency and energy consumption. 
 
Recently, parameter-efficient learning has been recognized as one potential solution to ameliorate the difficulties of adapting a large pre-trained language model, which aims to adapt a frozen pre-trained model by \textbf{only training some small additive modules} (e.g., residual adapter~\cite{houlsby2019parameter}, neural reprogramming~\cite{yang2021voice2series} and input prompt~\cite{chang2022exploration}). By integrating parameter-efficient learning, pre-trained language models~\cite{kenton2019bert} (PLMs) require less training time and computing resources to attain new state-of-the-art performance on different natural language processing tasks. 

In sum, how to advance parameter-efficient learning with existing English ASR models is one open topic for more voice applications without sufficient data sources like English. In this work, we propose \textbf{three specific designs of ASR model reprogramming} with Conformer~\cite{gulati2020conformer} based architectures for cross-lingual adaptation. As shown in Figure~\ref{fig:1:demo}, our proposed \textbf{C}onformer-based \textbf{A}SR \textbf{R}eprogramming (CAR) makes most of trainable neural architecture frozen (e.g., non-trainable) and only inserts few trainable modules for parameter-efficient model training.

\subsection{Parameter-Efficient Learning with Frozen ASR Models}
We review recent advances in parameter-efficient learning with frozen ASR models and justify its difference from a neural model reprogramming perspective.  Residual adapters~\cite{houlsby2019parameter} were initially introduced to vision domain applications~\cite{rebuffi2017learning} as a computationally efficient solution in contrast to full model-tuning. Houlsby \emph{et al.}~\cite{houlsby2019parameter} further advanced the design of residual adapters by developing a non-linear projection mechanism over latent features within frozen feature extractors (e.g., pre-trained transformer layers). Given the fact that acoustic feature encoders are standard components for ASR models, several recent works have demonstrated the effectiveness of applying residual adapters~\cite{houlsby2019parameter} for various speech applications, such as atypical speech~\cite{tomanek2021residual}, multilingual speech~\cite{kannan2019large}, and children's speech recognition~\cite{fan2022draft}. Meanwhile, there are related works studying how to build trainable parameters upon latent features from speaker adaptation~\cite{swietojanski2016learning} and latent space adversarial reprogramming literature. Bapna and Firat~\cite{bapna2019simple} have shown that residual adapters for ASR are much more effective than hidden unit modulation-based methods. However, the connections between each solution deserve more investigation, where model reprogramming literature~\cite{yang2021voice2series} has recently developed one first theoretical justification for the success of pre-trained speech model adaptation on population risk analysis.

\subsection{Neural Reprogramming: from Input to Latent Space}

The Neural reprogram (NR) method was first introduced in~\cite{elsayed2018adversarial} to re-purpose frozen pre-trained classifiers with a small amount of trainable input noise for out-of-domain image predictions. Recently, NR has been utilized for benchmarking sequence prediction tasks, such as classifying time series signals~\cite{yang2021voice2series} and spoken command~\cite{yen2021study}. NR mainly adds trainable parameters at its input level of a pre-trained model and thus enjoys federated advantages for distributed pre-trained models on-device. Meanwhile, NR has recently been introduced in latent space optimization~\cite{hambardzumyan2021warp} and demonstrates a state-of-the-art text classification performance. In the next section, we will introduce similar components of different existing parameter-efficient learning methods and provide a new design to summarize existing parameter-efficient learning techniques that empower cross-lingual speech recognition from a monolingual model.

\begin{figure}[t]
\begin{center}
\vspace{-2mm}
  \includegraphics[width=0.90\linewidth]{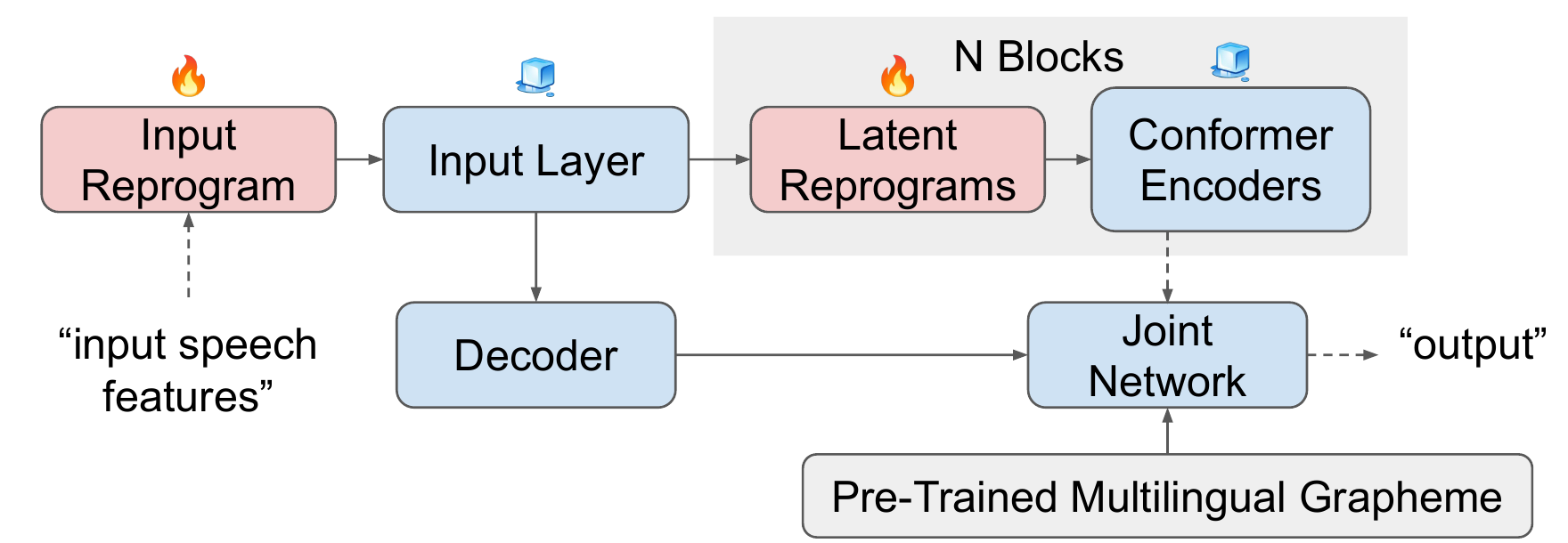}
\end{center}
\vspace{-0.2cm}
  \caption{A proposed design flow of conformer-based ASR reprogramming, which include a trainable input reprogramming for acoustic features and trainable feature reprograms for latent representations. 
  }
 \vspace{-0.4cm}
\label{fig:1:demo}
\end{figure}

\section{Neural Reprogramming for ASR Model}
\label{sec:repro}

Building upon the success of the aforementioned parameter-efficient learning techniques, our neural reprogramming has three major components: (1) input reprogramming, (2) latent space reprogramming, and (3) multilingual graphemes pre-training. As a short summary, (1) is associated with standard model reprogramming or input-prompting, (2) is related to the effectiveness of reprogramming and residual adapters, and (3) aims to resolve the existing challenges of cross-lingual learning on graphemes mismatching.

\subsection{Input Level Reprogramming}

Given a pre-trained neural network $\mathrm{M}$ with frozen model parameters $\Theta$, an input feature $x$, we can access its predicted output $y{'}=\mathrm{M}_{\Theta}(x)$  by feeding the input into the pre-trained model $\mathrm{M}_{\Theta}$. The goal of input level reprogramming is to find a trainable reprogramming function of $\mathcal{R}_\theta$ to minimize a prediction loss ($\mathcal{L}_{\text{error}}$) between $y{'}$ and its true label $\hat{y}$. In the previous speech model reprogramming studies~\cite{yang2021voice2series, yen2021study}, a trainable universal noise has been deployed for cross-domain adaptation, which is equivalent to our feature-independent reprogramming target $w_{\theta_2}$. However, in our empirical study, we find that only applying universal noise does not show a competitive performance, and we further introduce a feature-dependent trainable feature extractor ($\mathcal{H}_{\theta_1}$) in Eq.~(\ref{eq:argmin}). 

\begin{align}
\label{eq:argmin}
& \theta^*  = \arg \min_{\theta} \left \{  \mathcal{L}_{\text{error}}(\mathrm{M}_{\Theta}(\mathcal{R}_\theta(x)),\hat{y}) \right \}  \\
& \text{where}~~~\mathcal{R}_\theta(x)  = \underbrace{x}_{\text{original input}} + \underbrace{w_{\theta_2}}_{\text{feature-independent}} + \underbrace{\mathcal{H}_{\theta_1}(x)}_{\text{feature-dependent}} \nonumber
\end{align}

We have conducted an ablation study and select simple 1D-lightweight convolution~\cite{wu2018pay} followed by spatial attention~\cite{chen2017sca} encoding as best evaluating extractor setup for cross-lingual ASR. 

\subsection{Latent Space Reprogramming with Bridged Connections}
\begin{figure}[ht!]
        \centering
        \begin{subfigure}[b]{0.230\textwidth}
            \centering
            \includegraphics[width=\textwidth]{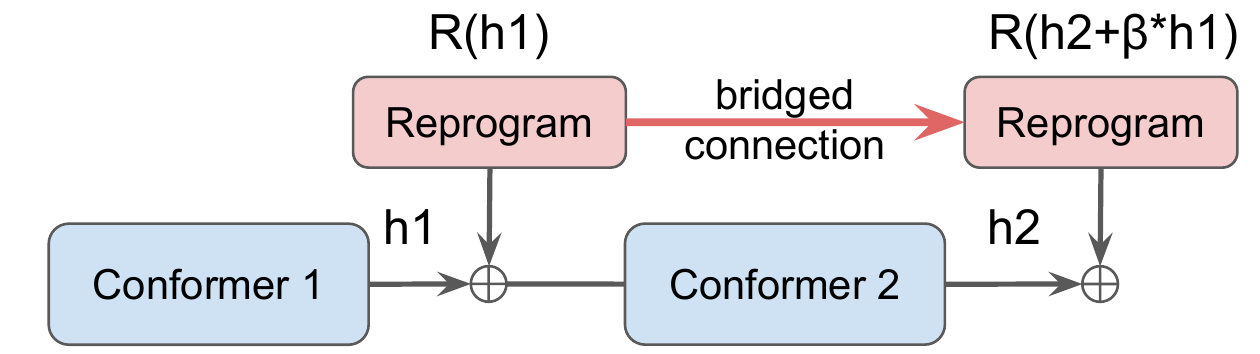}
            \caption[]%
            {{\small }}    
            \label{fig:2:bg:repro}
        \end{subfigure}
        \hfill
        \begin{subfigure}[b]{0.230\textwidth}  
            \centering 
            \includegraphics[width=\textwidth]{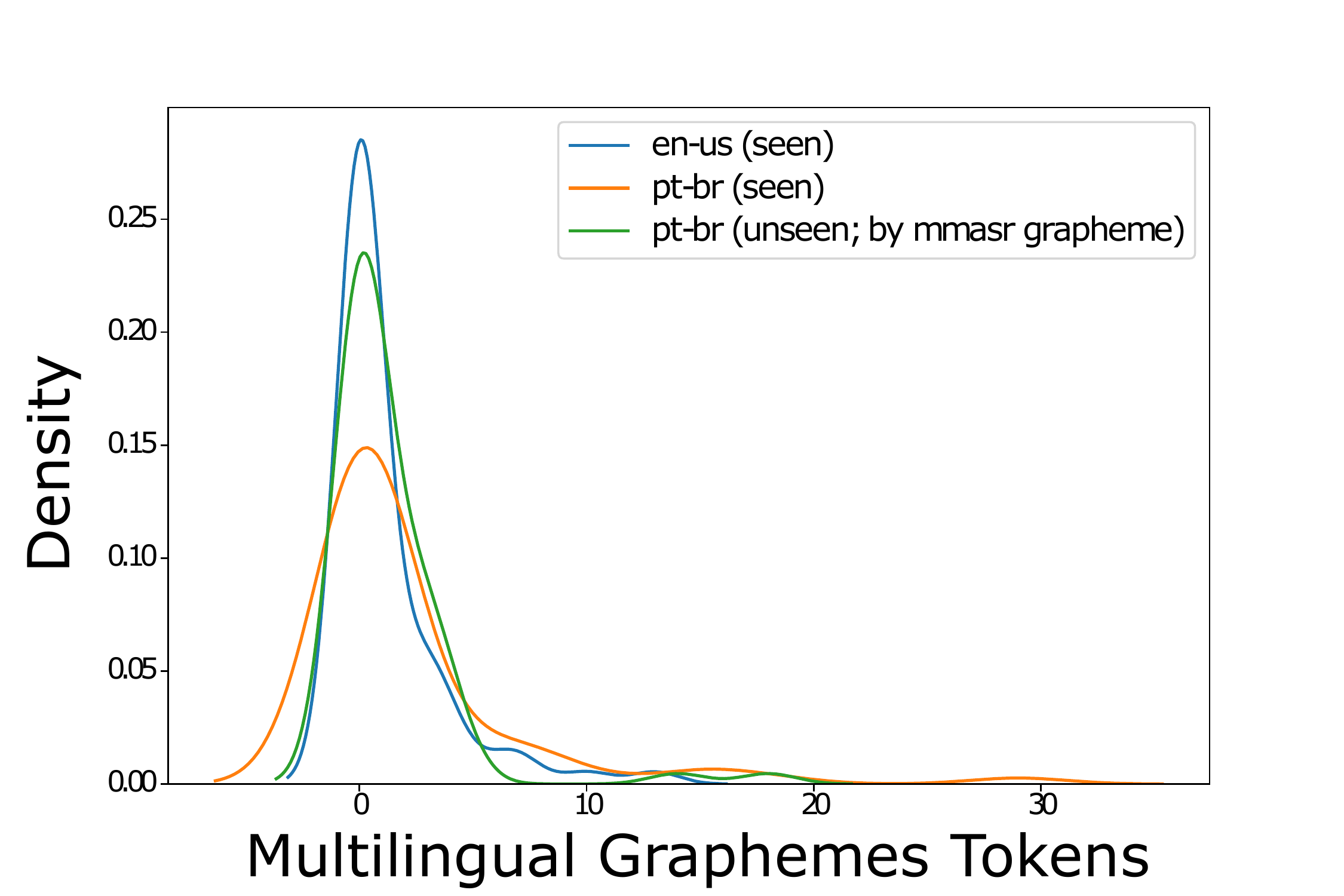}
            \caption[]%
            {{\small }}    
            
        \label{fig:2}
        \end{subfigure}
        \caption{(a) Bridged-connected reprogramming mechanism. (b) Multi-lingual grapheme distribution, where discriminative information is learned by English pre-training (\textcolor{teal}{green}) to improve generalization with a similar distribution of its ground-truth (\textcolor{orange}{orange}).}
        \vspace{-0.2cm}
\end{figure}

To further boost the performance of using a frozen Conformer-ASR model, we introduce extra trainable features in the latent space between each encoder. We call this baseline latent space reprogramming, which could be considered one inspiration of word-level reprogramming~\cite{hambardzumyan2021warp} and residual adapters~\cite{houlsby2019parameter} with optimization in the latent space. We further make a new design of ``bridged connection'' for latent space reprogramming to enhance additive feature learning on the frozen ASR model. Fig.~\ref{fig:2:bg:repro} shows how bridged-connected reprogramming blocks insert trainable feature between frozen conformer encoders. Given a $i$-th frozen conformer encoder as function $\mathcal{F}^i_{\Theta}$, we have latent feature $h^i$ from $i$-th conformer layer extracted from input $x$. The bridged-connection mechanism has been deployed for the following $(i+1)$-th conformer layer $\mathcal{F}^{i+1}_{\Theta}$ computing as the third term of Eq.~(\ref{eq:bg}) with a deterministic dropout rate ($\hat{\beta}=0.15$). In this work, we use the same reprogramming generator $\mathcal{R_{\theta}}$ to generate additive features for cross-lingual adaptation.

\begin{equation}
    \underbrace{\mathcal{F}^{i+1}_{\Theta}(h^i )}_{\text{frozen encoder}} \rightarrow \underbrace{\mathcal{F}^{i+1}_{\Theta}(\mathcal{R}_{\theta}(h^i) )}_{\text{latent reprogramming}}\rightarrow \underbrace{\mathcal{F}^{i+1}_{\Theta}(\mathcal{R}_{\theta}(h^{i}+ \hat{\beta} h^{i-1}) )}_{\text{bridged-connection reprogramming}}
    \label{eq:bg}
\end{equation}

\subsection{English Graphemes Pre-training for Multilingual Data}
\label{sec:en:gh}
To effectively adapt large-scale pre-trained English ASR models to different language recognition (e.g., English to French), previous research efforts~\cite{zhao2021addressing} suggest a solution by replacing the last prediction layer of ASR. However, we investigate that deploying ``multilingual graphemes'' is even more effective than replacing the final prediction head directly. As shown in Fig~\ref{fig:2}, we find that multilingual graphemes with English pre-training also learn some discriminative information to feed unseen utterances (e.g., Portuguese). Table~\ref{tab:en-us} shows an investigation of the importance of multilingual grapheme for attaining a lower classification error rate. A unified multilingual grapheme set with $80$ tokens~\cite{li2021scaling} has been selected as output vocabulary of our ASR systems. Noted that we have investigated that the performance gap between using monolingual (e.g., English-only) and multilingual graphemes is with relatively slight degradation ($\pm$0.11\%). Tuning conformer models with extra dense layer~(F0b \& F1b), also called as linear probe, does not outperform fine-tuning (F0 \& F1) baselines, with $4$k more trainable parameters.

\begin{table}[ht!]
\centering
\caption{The importance of multilingual grapheme pre-training on English (denoted as gh$^{\text{multi}}_{\text{en}}$) with pre-trained Conformer RNN-T.}
\label{tab:en-us}
\begin{tabular}{|l|c|}
\hline
Setup (from a en-ASR) & WER \\ \hline \hline
\textbf{F}0: \textbf{F}ine-tuning  all (w/ gh$^{\text{multi}}_{\text{en}}$) & \textbf{10.5} \\ \hline \hline
F0a: F0 w/o loading gh$^{\text{multi}}_{\text{en}}$ & 13.4 \\ \hline
F0b: F0 w/ extra dense layer & 12.6 \\ \hline \hline
\textbf{F}1: \textbf{F}ine-tuning  last conformer (w/ gh$^{\text{multi}}_{\text{en}}$) & \textbf{18.2} \\ \hline \hline
F1a: F1 w/o loading gh$^{\text{multi}}_{\text{en}}$ & 49.2 \\ \hline
F1b: F1 w/ extra dense layer & 21.1 \\ \hline
\end{tabular}
\end{table}
\vspace{-4mm}

\section{End-to-End Conformer-based ASR Systems}
\label{sec:sys}
This section presents our E2E-ASR system and introduces how model reprogramming could outperform other parameter-efficient learning solutions. We study the cross-lingual adaption tasks in two E2E-ASR systems: (1) training with a supervised training loss only; (2) a joint supervised and unsupervised loss (e.g., with w2v-Bert). Our multilingual ASR model is a Conformer-based RNN-T architecture. We conduct our parameter-efficient learning experiments mainly on the RNN-T, where the findings could generalize to the other encoder-decoder ASR pre-training for future studies. 
\subsection{System 1: Supervised Training for Conformer-based ASR}
\label{sec:sup:3:1}
We select our pre-training Conformer RNN-T system based on the previous work~\cite{li2021scaling}, which attains competitive multilingual recognition performance with only supervised training objective. The Conformer RNN-T includes an encoder network, a decoder network, and a joint prediction network. For the encoder, we deploy full-context Conformer layers, including an input projection layer, a relative position embedding layer followed by a stack of 17 Conformer layers. 

Similar to~\cite{li2021scaling} the stacked conformer layers can be categorized into three encoder blocks. The first block consists of $4$ Conformer layers and a time stacking layer for computing time reduction; the second block consist of a single Conformer layer (the fifth conformer layer) and a projection layer to map the current feature dimension back to its original. The remaining $12$ Conformer layers comprise the third encoder block. We use the existing convolution module in Lingvo~\cite{shen2019lingvo} toolkit to support relative positional information and group normalization in each Conformer layer. A 2-layer of 
unidirectional LSTM network have been used as decoder. The supervised Conformer ASR has 270M of trainable parameter in total. 

To empower parameter-efficient learning, we firstly train our model with $44.6$k hours of English data from MLS~\cite{pratap2020mls}. The Conformer ASR attains a competitive $5.5$\% WER with the aforementioned multilingual graphemes discussed in Sec.~\ref{sec:en:gh}. We then make most of the most parameter non-trainable (e.g., frozen as shown in Fig.~\ref{fig:1:demo}) and only train the reprogram layers loading from a pre-training English Conformer RNN-T. We make an ablation study inspired by the existing residual adapters research to justify our design. 

\subsection{System 2: Supervised ASR with Self-Supervised Losses}
\label{sec:unp:3:2}
We further investigate that advanced supervised ASR with unsupervised pre-training could be trained under only small trainable parameter budgets to attain high cross-lingual recognition performance. We utilize the previous joint unsupervised and supervised training (JUST) ASR model~\cite{bai2022joint} to combine the supervised RNN-T loss and the self-supervised learning (SSL) (i) contrastive and (ii) masked language modeling (MLM) losses. In the JUST ASR model, both self-supervised modules of the Contrastive net and MLM net are a stack of 
8 and 16 standard conformer layers. Following the basic setup in ~\cite{bai2022joint}, we freeze those conformer layers and inserted the proposed to reprogram layer with bridged connections in between during the model training. The total of pre-trained JUST ASR model is defined as $\mathcal{L}_{\text{JUST}}=\mathcal{L}_{\text{
rnn-t}}+
\gamma(\mathcal{L}_{\text{c}}+\mathcal{L}_{\text{mlm}}+\alpha\mathcal{L}_{\text{div}})$, where $\mathcal{L}_{\text{
rnn-t}}$ is the supervised loss as the discussion in Sec~\ref{sec:sup:3:1}, $\gamma$ is the joint coefficient for SSL losses (set to be 0.01), $\mathcal{L}_{\text{c}}$ is the loss of Contrastive net, $\mathcal{L}_{\text{mlm}}$ is the loss of MLM net, and entropy-based diversity loss ($\mathcal{L}_{\text{div}}$) is used for code-book related optimization with $\alpha=0.1$ followed the similar implantation in previous wav2vec2 study~\cite{baevski2020wav2vec}.

\section{Experiment and Results}
\label{sec:exp}
In this section, we introduce basic setups and proposed parameter-efficient reprogramming in three cross-lingual ASR tasks. Noted that the Para. in Tab~\ref{tab:3:major} and Tab~\ref{tab:5:just:repro} means trainable model parameters. 

\subsection{Setup and Parameter-Efficient Architectures}
\textbf{Dataset:} we conduct our experiments on the popular Multilingual Librispeech~\cite{pratap2020mls} (MLS) benchmarks on the eight languages, including (1) Germanic languages of English (en) with 44.6k hours, German (de) with 1.96k hours and Dutch (nl) with 1.55k hours; (2) Romance languages of French (fr) with 1.1k hours, Spanish (es) with 0.9k hours, Italian (it) 0.2k hours, Portuguese (pt) with 0.16k hours; and (3) west slavic languages of Polish (pl) with 0.1k hours, which are one widely studied and standard evaluation set in the MLS. We use the official test set of MLS to report test word-error-rate (WER) performance for each setup. 

\textbf{Features and ASR systems:} Log-Mel filterbank features (80-dims) are used as inputs extracted from MLS acoustic utterances (e.g., 10-20 seconds long.) The encoder layer of supervised Conformer ASR (Study 1) is introduced in Sec~\ref{sec:sup:3:1}, also referred to in the details presented in~\cite{li2021scaling}. For JUST-based Conformer ASR, both Contrastive net and MLM net have 1024 hidden units with 8 attention heads. The sample masking ratio for MLM is $6.5$\% with a codebook size of $1024$. Please refer to~\cite{bai2022joint} for more JUST-related details. 
 
\textbf{Trainable Parameter Budget and Hyperparameters:} we carefully control reprogramming and the other auxiliary tuning baselines under a similar parameters and report its best setup, where recent memory efficient ASR study~\cite{venkatesh2021memory} has studied the same scale of $\sim$10M to 30M trainable parameters. Our final goal is to find out a best architecture to adapt large-scale frozen pre-training ASR models. We train supervised Conformer-ASR with batch size 1024 on 64 TPUs and JUST-ASR with batch size $1024$ on $256$ TPUs. For gradient training, Adam optimizer is suggested to be $\beta_1$ = $0.9$, $\beta$ = $0.98$. For JUST training, we use a global learning rate scheduler as described in~\cite{bai2022joint}. For residual adapter, we utilize the benchmark design from~\cite{houlsby2019parameter, tomanek2021residual} with a latent dimension of $256$ after ablations.

\begin{table}[]
\centering
\caption{\textbf{C}onformer-\textbf{A}SR \textbf{R}eprogramming (CAR) for $\mathrm{Study}_1$.}
\label{tab:3:major}
\begin{tabular}{|l|c|c|}
\hline
Setup  & Avg. WER & Para$_\textit{train}$ \\ \hline
\textbf{B}0: \textbf{B}aseline frozen en-Conformer & 92.1 & 0 \\ 
B1: Conformer training from scratch & 10.7 & 270M \\ 
B2: Wav2Letter~\cite{pratap2020mls} & 11.8 & 100M \\
B3: XLSR-53 (w/ external data)~\cite{conneau2020unsupervised} & 10.6 & 300M \\ \hline
\textbf{F}0: \textbf{F}ine-tuning from en & 10.5 & 270M \\ \hline
F1: Fine-tuning last conformer & 18.2 & 13M \\ 
F2: Adding an extra conformer & 34.9 & 13M \\ 
F3: Residual adapters from en & 13.5 $\pm$ 0.2 & 11M \\ 
F4: Fine-tuning decoder & 20.9 & 20M \\ 
F5: Bias-terms fine-tuning (BitFit~\cite{zaken2022bitfit})& 33.0 & 0.2M \\ \hline
CA\textbf{R}1: Attention-\textbf{R}eprogramming  & 12.6 $\pm$ 0.4 & 11M \\
CAR2: Conv-Reprogramming& 13.0 $\pm$ 0.2 & 11M \\
CAR3: CAR1 + Bridged Connection &\textbf{11.9} $\pm$ 0.3 & 11M \\ \hline
\end{tabular}
\end{table}
\vspace{-2mm}
\subsection{Cross-Lingual Speech Recognition Results}
We evaluate different architecture and their performance with frozen supervised conformer-ASR (System 1 in Sec~\ref{sec:sup:3:1}) for cross-lingual recognition results. After finding one best parameter-efficient learning setup, we further evaluate if the best setup could further train a supervised conformer model with SSL losses (System 2 in Sec~\ref{sec:unp:3:2}).

\textbf{$\mathrm{Study}_1$: Monolingual ASR from English Pre-training}\\
We first investigate the monolingual ASR result. For evaluation, we independently train on the seven non-en data (es, it, pt, fr, de, nl, pl) from scratch and report its average WER after 10 runs. Based on the results, direct using a frozen Conformer-ASR pre-trained from en should yield WER above $90$\%. As shown in the fifth (F0) to ninth row (F4) in Table~\ref{tab:3:major}, the residual adapter-based method attains a lower WER of $13.5$\% averaged in the seven languages. Tuning the conformer layer itself does not provide good performance as the residual adapter, whether (F1) directly tuning the last layer of the Conformer encoder or (F2) training an extra conformer layer followed the encoder. Noted that we have also conducted experiments on one-layer by one-layer tuning but fine-tuning the last conformer outperforms tuning the other single conformer layer. Meanwhile, we also found that the boosted performance under the frozen ASR scheme is \textbf{mainly coming from tuning the \textit{encoder}}, where directly tuning the \textit{decoder} (F4) of RNN-T produces a WER above $20$\%. Similarly, \textit{bias-terms} only fine-tuning (BitFit)~\cite{zaken2022bitfit} (F5) shows a $33$\% WER, which have not yet performed competitively in RNN-T based ASR modeling, according to our empirical evaluation on the MLS. 

Next, we study impacts of architectural differences under the proposed scheme of conformer-ASR reprogramming (CAR). We first investigate the difference between the additive feature generators. We find that the 1D-convolution-based feature extractor had a similar performance to the residual adapters. We use spatial attention to encode input and use 2D-convolution over the acoustic feature to reduce trainable parameters. This setup shows a boosted performance compared to its convolution-based ablation and outperforms residual adapters by  $3.7$\% WER relative. We add the bridged connection to the best attention-based reprogramming setup, which reduces by $5.6$\% WER relative. Since the bridged connection provided an additional gradient path (between each reprogram layer) apart from its backbone model, we achieve a \textbf{reduced} $10.1$\% computing time during the model training compared to residual adapters~\cite{kannan2019large}. We select attention-based reprogramming with bridged connection as our best setup (denoted as reprogram$_\text{CAR3}$) to investigate other adaptation properties further. 

\textbf{$\mathrm{Study}_2$: Tuning from Multilingual ASR Pre-trainings}\\
Whether ``multilingual`` pre-training could serve as a better format of pre-trained ASR backbone is one open question for cross-lingual ASR. In Table~\ref{tab:4:multi}, we aim to tackle this question by carefully controlling similar total training hours (44 to 45k hours) for the same supervised Conformer-ASR with different extra mixed languages of ``en+fr'' or ``en+es.'' We then report its test WER taking the average on five unseen languages of de, nl, it, pt, and pl. Interestingly, both multilingual ASR backbones demonstrate better performance than en-only Conformer-ASR by $1.6$ to $0.5$\% absolute WER.

\begin{table}[ht!]
\centering
\caption{\textbf{M}ulti-to-mono ASR ($\mathrm{Study}_2$) under similar total hours.}
\label{tab:4:multi}
\begin{tabular}{|l|c|c|c|}
\hline
Setup \textbackslash{} Pre-Training Languages & en & en+fr & en+es \\ \hline
\textbf{M}0: Fine-tuning all & 11.3 & \textbf{9.7} & 10.1 \\ \hline
M1: Residual Adapter & 13.9 & \textbf{12.5} & 12.9 \\ 
M2: Reprogramming$_\text{CAR3}$ & 12.8 & \textbf{11.8} & 12.3 \\ \hline \hline
Total training hours & 44.6k & 45.6k & \textbf{45.7k} \\ \hline
Covered Graphemes out of 80 & 29 & \textbf{51} & 42 \\ \hline
\end{tabular}
\end{table}

\textbf{$\mathrm{Study}_3$: Tuning from ASR with Self-Supervised Losses}\\
Since JUST-ASR contains w2v-BERT pre-training from Libri-Light~\cite{kahn2020libri} with $60$k hours of unannotated en-speech, we report its WER performance on en and the seven languages for recognizing multilingual speech at the same time followed the setup in~\cite{bai2022joint}. For our parameter-efficient study, we only train (i) the proposed bridged-reprogram layers and (ii) its original decoder layer to update unsupervised losses discussed in Sec~\ref{sec:unp:3:2} in a few-shot learning setup with 100k only step updates in TPUs. As an extra finding, either reprogramming or using adapters with fully unsupervised w2v-Bert~\cite{chung2021w2v} would only generate poor WER  over $70\%$, which indicates the importance of supervised loss. As shown in Table~\ref{tab:5:just:repro}, we notice that fine-tuning together with reprogramming modules (J1) could reduce both en (-$2.7$\%) and unseen (-$11.6$\%) WER relatives from FT JUST (J0). We have confirmed that the performance gains of parameter-efficient solutions, J2 and J3, \textbf{mainly come from the reprogramming or adapters} compared to FT decoder-only (J4).

\begin{table}[ht!]
\centering
\caption{Reprogramming JUST for Multilingual ASR ($\mathrm{Study}_3$)} 
\label{tab:5:just:repro}
\begin{tabular}{|l|c|c|c|}
\hline
\multicolumn{1}{|c|}{Setup} & en & 7-lang & Para. \\ \hline 
\textbf{J}0: Fine-tuning (FT) \textbf{J}UST~\cite{bai2022joint} & 7.5 & 9.5 & 660M \\ \hline
J1: FT JUST + Reprogram$_\text{CAR3}$ & \textbf{7.3} & \textbf{8.4} & 710M \\ 
J2: Reprogram$_\text{CAR3}$ + FT decoder & \textbf{8.4} &\textbf{9.7} & 45M \\ 
J3: Adapter + FT decoder & 8.9 & 10.2 & 45M \\ \hline
J4: FT decoder only & 17.4 & 22.0 & 20M \\  \hline
\end{tabular}
\end{table}
\vspace{2mm}

\textbf{Limitation and Additional Discussion}
When the current Conformer-ASR reprogramming performs well in different cross-lingual evaluations, we want to remind audiences that one challenge for adapting languages with large grapheme tokens (e.g., Mandarin and Japanese). One potential solution is to use dictionary learning to composite phoneme tokens, and we leave it for future works. Note that the current prompt-tuning~\cite{chang2022exploration} for speech mainly covers cross-tasks adaption and yet for ASR studies, where the input-reprogramming could be one similar approach along this direction.  

\section{Conclusion}
\label{sec:conclusion}
This work introduces a novel parameter-efficient learning solution for cross-lingual ASR. Our proposed model reprogramming module leverage upon its design on adding trainable attributes on both input and latent space, which further constructs a light-weighted solution to adapter large-scale pre-trained ASR models. For the supervised ASR model, we only require 11M ($4.8$\% of its full pre-trained model) trainable parameters to achieve $11.9$\% WER cross seven languages in MLS benchmark. Model reprogramming shows competitive $8.4$\% and $9.7$\% WERs combining SSL setups, such as transferring a 660M pre-trained model with only $6.8$\% of its original model parameters. Our proposed method and new findings on cross-lingual recognition could be considered as one preliminary pathway to designing a  large ``foundation speech model'' for future studies.

\clearpage
\footnotesize
\bibliographystyle{IEEEtran}

\bibliography{refs}

\end{document}